\documentclass[structabstract]{aa}  
\usepackage{graphicx}
\usepackage{txfonts}
 \usepackage{lscape}
 \usepackage{colortbl}
 \usepackage{natbib}
 \usepackage{verbatim} 
  \usepackage{comment} 
\usepackage{url}
 \bibpunct{(}{)}{;}{a}{}{,}
\begin{document}
   \title{ 
The role of environment in the morphological transformation of galaxies  in 9 rich intermediate redshift clusters. }
\titlerunning{Wide field study of $z\sim0.5$ clusters}

  \author{M. Huertas-Company 
          \inst{1,2}
          \and
          G. Foex
          \inst{3}
          \and
          G. Soucail
          \inst{3}
          \and
          R. Pell\'o
          \inst{3}
          }

   \institute{ESO, 
   		Alonso de Cordova 3107 - Casilla 19001 - Vitacura -Santiago, Chile	 \email{mhuertas@eso.org}
           \and
           LESIA - Paris Observatory, 
   5 Place Jules Janssen, 92195 Meudon, France
       \and
           Laboratoire d'Astrophysique de Toulouse-Tarbes, CNRS and Universit\'e de 
Toulouse, 14 Av.\ E.~Belin, F--31400 Toulouse, France
            }

   \date{Received September 15, 1996; accepted March 16, 1997}

 
  \abstract
   {Rich clusters offer a unique laboratory for studying the effects of local environment on the morphological transformation of galaxies moving from the blue star-forming cloud to the red passive sequence. Due to the high-density, any environmental process should be more pronounced there compared to the field population. }
   {Ideally, we would like to reconstruct the evolution of a single, hypothetical representative galaxy as it enters the cluster. For that purpose wide-field imaging is crucial to probe a wide range of densities and environments (from the core to the outskirts) and isolate this way, the different physical processes which are responsible of the migration from the blue-cloud to the red-sequence.}
   { We analyze a sample of 9 massive clusters at $0.4<z<0.6$ observed with MegaCam in 4 photometric bands (g,r,i,z) from the core to a radius of $5$ Mpc ($\sim4000$ galaxies). Galaxy cluster candidates are selected using photometric redshifts computed with HyperZ. Morphologies are estimated with \textsc{galSVM} in two broad morphological types (early-type and late-type). We examine the morphological composition of the red-sequence and the blue-cloud and study the relations between galaxies and their environment through the morphology-density relations ($T-\Sigma$) and the morphology-radius relation ($T-R$) in a mass limited sample ($log(M/M_{\odot})>9.5$). }
   {We find that the red sequence is already in place at $z\sim0.5$ and it is mainly composed of very massive ($log(M/M_{\odot})>11.3$) early-type galaxies. These massive galaxies seem to be already formed when they enter the cluster, probably in infalling groups,  since the fraction remains constant with the cluster radius. Their presence in the cluster center could be explained by a segregation effect reflecting an early assembly history. Any evolution that takes place in the galaxy cluster population occurs therefore at lower masses ($10.3<log(M/M_{\odot})<11.3$). For these galaxies, the evolution, is mainly driven by galaxy-galaxy interactions in the outskirts as revealed by the $T-\Sigma$ relation. Finally, the majority of less massive galaxies ($9.5<log(M/M_{\odot})<10.3$) are late-type galaxies at all locations, suggesting that they have not started the morphological transformation yet even if this low mass bin might be affected by incompleteness. }
   {}
   \keywords{Galaxies: clusters: general -- Galaxies: evolution -- Galaxies: high-redshift}

   \maketitle
%

\section{Introduction}
Studies of galaxy properties and their dependences on environment are key to understanding the mechanisms that are responsible for galaxy evolution since they constitute a natural test for cosmological theories of galaxy formation. Evidence has accumulated in recent years that environmental processes affect both the star formation and morphological characteristics of galaxies, the endpoint being the present-day morphology-density relation \citep{Dressler80, Treu03}. 

In particular, two fundamental observational results deal with the star formation history and the color-morphology distribution of field galaxies: a) the downsizing (e.g. \citealp{Cowie96, Bundy05}), i.e. a stronger and earlier decline of star formation in massive galaxies compared to a slower decrease of the star formation rate of the present blue/late-type galaxies and b) the bimodality (e.g. \citealp{Strateva01, Baldry04}) of the galaxy distribution in the color-mass space, i.e. a marked segregation in color between the blue cloud (low mass, star-forming late-type galaxies) and the red-sequence (massive, red, passive early-type galaxies). 

Therefore, the scenario which seems to emerge is that galaxies move from the blue population to the red-sequence at rates and cosmic epochs that depend very strongly on their mass and environment. 
However, the nature and timescale of the relevant mechanisms that lead this mass-dependent transition are still unclear. We need processes to quench the star formation above a mass threshold such as gas loss caused by AGN feedback and/or supernova feedback (e.g. \citealp{DiMatteo05, Menci06, Neistein06}), consumption of all available gas in the wake of violent merging (e.g. \citealp{Faber07}), ram-pressure stripping/harassment in clusters that turns off efficient cold gas accretion. A combination of these processes are normally required in semi-analytical models to ensure that quenching is efficient, permanent, and mass-dependent \citep{delucia06}.


From that point of view, rich clusters offer a unique laboratory for studying the effects of local environment on the transition between the blue cloud and the red sequence (e.g. \citealp{poggianti06}). Their cores have the highest volume density of galaxies in the universe so that any environmental dependence of galaxy formation or evolution processes should be more pronounced there, when contrasted with studies of the field population.
 Ideally, we would like to make a significant step forward and reconstruct the complete evolution of a single, hypothetical representative galaxy as it enters the cluster potential and interacts with the ICM. How do star-formation and morphology evolve as a galaxy enters the cluster? How does it move form the blue-cloud to the red-sequence? To answer these questions it becomes necessary to assign timescales to the phenomena observed and to connect afterwards clusters at different redshifts into a temporal sequence. The various processes that can take place can be classified in three broad headings (see \citealp{Treu03} for a nice review): 1) galaxy-ICM interactions, which refers to the interaction of a cluster galaxy with the gaseous component of the cluster, 2) galaxy-cluster gravitational interactions, including tidal and related dynamical processes and 3) smaller scale galaxy-galaxy interactions. 
  
During the past years, much progress have come from the connection of Hubble Space Telescope (HST) imaging with deep ground-based spectroscopy. Several HST-based studies have demonstrated that the blue galaxies seen at $z>0.3$ are star-forming galaxies which are absent in present day clusters. (e.g. \citealp{Dressler94, Stanford95}) It has been proposed that these are recent arrivals in the cluster, probably observed before the removal of their gas reservoir \citep{delucia07}. 

The main problem of HST imaging is however its small field of view, which makes it very time consuming to obtain a wide view of the cluster from the inner core to the outskirts. That is the main reason why most of these previous works focus on the inner 0.5 Mpc of several clusters which is lower than the virial radius ($\sim2$Mpc).
However probing the outskirts of clusters is essential to discriminate between the different
processes that can induce morphological transformation of a galaxy entering the cluster.
\cite{Treu03} have shown indeed that depending on the analyzed zone it is possible to isolate
one process from another. Thanks to a wide survey on a single target, they were able to show
that the most likely processes responsible for the mild gradient in the morphological mix outside the virial radius are galaxy-galaxy interactions.
This later affirmation has been confirmed in more recent works. Using a volume-limited sample, \cite{Patel09} argued that the evolution in the galaxies entering a cluster at $z\sim0.8$ is mainly driven by the local environment. In this last work, the authors also showed an interesting trend: the dependence of galaxy evolution in clusters with the stellar mass. They indeed identify the same downsizing effect as seen in the field \citep{Cowie96, Bundy05}, with massive galaxies having their star forming rate decline faster.  
  
At this point, it becomes interesting to have a wide view of a statistically representative
sample of clusters at high redshift with robust stellar mass estimates to probe the physical
mechanisms responsible of the mass-dependent transition from the blue-cloud to the red-sequence. 
MegaCam, the wide field imager installed at the Canada France Hawaii Telescope can probe in a single pointing a cluster at $z\sim0.5$ from the core to $\sim12$ Mpc radius. Problems come now from angular resolution, since determining morphologies of galaxies from this ground-based images is not trivial. \cite{Huertas-Company08a} describes a method to estimate robust morphologies on poorly resolved data. In that work, the authors were able to prove that, when applied to seeing-limited data, the method can provide morphological information in two broad classes, with an accuracy close to that obtained with space data.

In the present paper we use this method combined with accurate photometric redshift estimates
in 9 rich clusters at intermediate redshifts ($0.4<z<0.6$). We show that we are able to isolate galaxy cluster members and to estimate the morphologies of $\sim4000$ galaxies with a fair degree of reliability. We then connect them to the cluster environment in three mass ranges to probe the mass-dependent transition from the blue-cloud to the red-sequence.

The paper is organized as follows: in \S~\ref{sec:data_desc} we introduce the observational data. In \S~\ref{sec:gal_sel} we describe our method for selecting member galaxies based on photometric redshifts. We combine the photometric redshift estimate with the shape of the redshift probability function ($P(z)$) function to establish a robust criteria for selecting galaxies. The physical properties computed are described in \S~\ref{sec:phys_prop}. We specially focus on the morphology determination and on the accuracy quantification. We then describe and discuss our main results in \S~\ref{sec:results}.

Throughout the paper, we assume that the Hubble constant, the matter density, and the
cosmological constant are $H_0=70$ km s$^{-1}$ Mpc$^{-1}$, $\Omega_m=0.3$ and $\Omega_\Lambda=0.7$ respectively.
 
\section{Data description}
\label{sec:data_desc}
The data used in this paper are part of an imaging survey of X-ray
clusters at intermediate redshift ($z \sim 0.5$), undertaken to
study the mass distribution in these clusters and to compare the X-ray
distribution of the intra-cluster gas and the dark matter distribution
traced by weak lensing. The sample has been selected to be representative
of the rich cluster population at intermediate redshift and is part of the full
and unbiased sample observed with XMM-Newton as part as a Large Programme
(PI: M. Arnaud). As the main goal of the optical follow-up was to study
in details the mass distribution by weak lensing techniques, only the
hottest and most massive clusters of the sample were selected. This
ended with a sub-sample of 10 clusters, 9 of which are studied in this
paper. The last one, RXJ1347.5-1145 was discarded because it was observed
in different conditions at CFHT and so would reduce the homogeneity of the
study. The 9 resulting clusters were observed with the wide-field imager
MegaCam installed at CFHT, and in 4 photometric bands ($g',r',i',z'$)
and with integration times of 1600 sec., 7200 sec., 1200 sec. and 1800
sec. respectively. We asked for a longer exposture time in $r'$ because
these data are dedicated to the weak lensing measurement of the mass
distribution in the clusters, while the color information is restricted
to the study of the cluster population. We will therefore use $r'$ band data for the morphological analysis as well (\S~\ref{sec:morpho}). The total field of view covered
by the mosaic is one square degree, with a pixel scale of 0.187\arcsec ($\sim1.1$ Kpc at $z=0.5$). At the average cluster redshift ($z\sim0.5$), this corresponds to a
radius up to $\sim 12$ Mpc, well above the virial radius ($\sim1-2$ Mpc, see table~\ref{tbl:clusters_prop}). Consequently,
the outskirts of the clusters as well as they global environment can
also be probed. The basic properties of the clusters are summarized in
table~\ref{tbl:clusters_prop} and more details can be found in Foex et
al. (in preparation).

\begin{figure*} 
 \centering 
  \resizebox{\hsize}{!}{\includegraphics{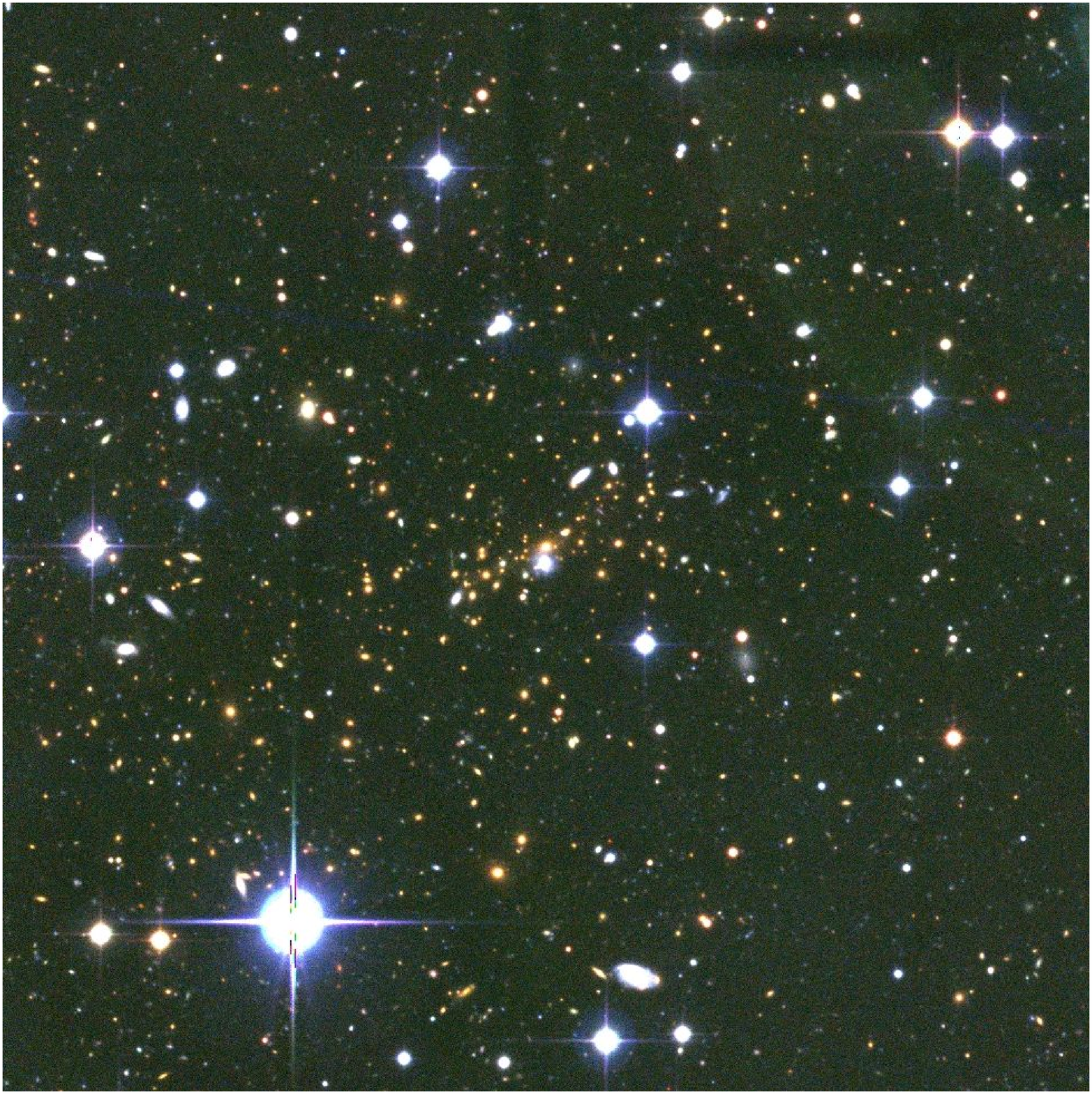}}
 \caption{Color image of a $6^{'}\times6^{'}$ ($\sim$2.5Mpc$\times$2.5Mpc) region around the center of MS 0451.5-0305 . The image has been built with MegaCam g, r and i filters.  }
 \label{fig:color_image}
 \end{figure*}

Data reduction was done in a standard way in the Terapix
center\footnote{http://www.terapix.fr} or using the specific tools
developped by Terapix (astrometry, stacking and photometric
calibration). The photometric catalogs were built using
\textsc{SExtractor} \citep{Bertin96} in dual mode for multi-color
photometry. As illustration, we show in Fig.~\ref{fig:color_image},
a color composite image of one of the clusters. On average, the
image quality on the final $r'$ images is $0.7\pm 0.1\arcsec$ with
some variation from cluster to cluster. The completeness magnitude
of the photometric catalogues is about $r'=25$, $g'=25$, $i'=24$ and
$z'=23.5$. For the details of the photometric survey of these clusters,
we refer to Foex et al. (2009, in preparation). In this paper, we
restrict the analysis to objects with $18<r<23$ lying at a radius lower
than $5$ Mpc. As we will show in section~\S \ref{sec:phys_prop} this is
equivalent to a mass cut at $\log(M_{*}/M_\odot)>9.5$.  The magnitude
cut is motivated by the reliability of the morphological classification
(\S~\ref{sec:morpho}) while the distance cut is done to reduce the field
contaminations (\S~\ref{sec:gal_sel}).

\begin{table*}

\caption{Main properties of clusters in our sample. [1] \cite{Bohringer00}, [2] \cite{Sadat04}, [3] Foex et al. 2009, in prep. $r_{200}$ are computed through a 1D weak-lensing analysis, i.e. fit of the radial shear profile and are defined as the radius at which the density is 200 times the critical density of the Universe at the cluster redshift.}
\begin{tabular}{lrrrrrr}
\hline\hline\noalign{\smallskip}
ID &$\alpha$ (J2000) & $\delta$ (J2000) & z & $L_X$ & $r_{200}$  \\ 
& & & & ($10^{-37}$ $W$) &  (Mpc)\\  
\noalign{\smallskip}\hline

MS 0015.9+1609  & 00:18:33.33  & +16:26:44.6 & 0.54 [2] & 15.5 [2]  & 2.11 [3]\\ 
MS 0451.5-0305  & 04:54:10.796 & -03:00:53.0 & 0.54 [2] & 23.7 [2] & 1.47 [3]\\ 
MS 1621.5+2640  & 16:23:35.6277 & +26:34:09.49 & 0.43 [2] & 9.47 [2]  & 1.5 [3]\\ 
MS1241.5+1710  & 12:44:01.4 & +16:53:42 & 0.55 [2] & 1.73 [2] & 1.7 [3]\\ 
RXC J0856.1+3756   & 08:56:08.5 &+37:56:52  & 0.41 [1] & 15.76 [1] & 1.59 [3]\\ 
RX J1003.0+3757  & 10:03:02.74 & +37:57:51.5 & 0.41 [1] & 8.29 [1] & 1.48 [3]\\ 
RX J1120+4318   & 11:20:07.5 & +43:18:05 & 0.60 [2] & 6.07 [2] &  1.3 [3]\\ 
RXC J1206.2-0848  & 12:06:13.1  & -08:47:43 & 0.44 & -- &  2.0 [3]\\ 
RXC J2228.6+2036  & 22:28:37.1 & +20:36:31& 0.41 [1] & 29.62 [1] &  1.62 [3]\\ 
\noalign{\smallskip}\hline
\noalign{\smallskip}\hline\noalign{\smallskip}
\label{tbl:clusters_prop}
\end{tabular}
\end{table*}

\begin{figure*} 
 \centering 
\resizebox{\hsize}{!}{\includegraphics{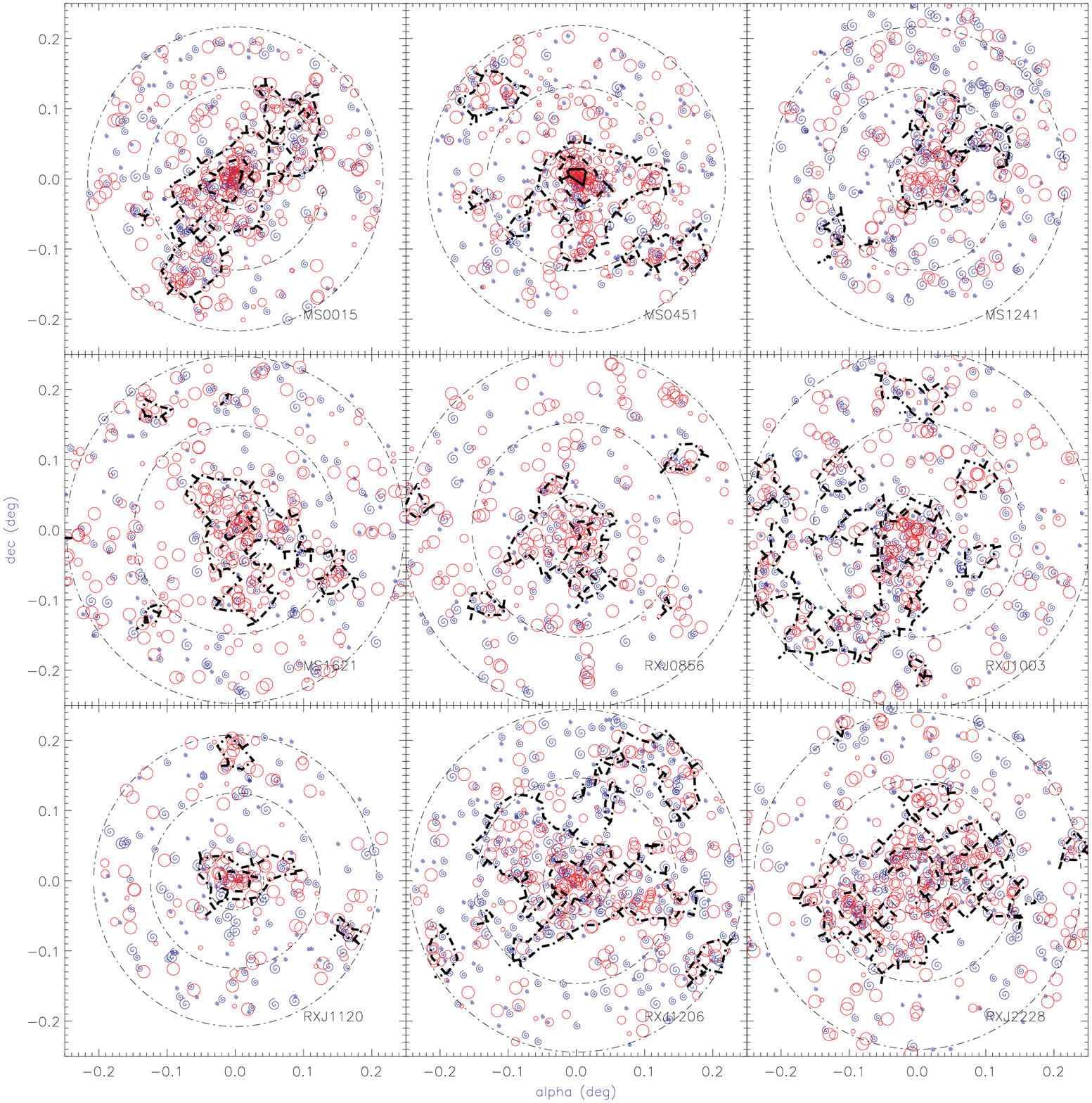}}
 \caption{Spatial distributions of galaxies in the different clusters. Red circles indicate early-type objects and blue spiral symbols late-type galaxies. The symbol sizes code three different mass ranges; big: $log(M/M_\odot)>11.3$, intermediate: $10.3<log(M/M_\odot)<11.3$ and small: $9.5<log(M/M_\odot)<10.3$.  Black circles show physical radii: 1, 2 and 5 Mpc. Contour lines indicate density levels: $D>2Mpc^{-2}$ (solid line), $D>1.5Mpc^{-2}$ (dashed line), $D>1Mpc^{-2}$ (dashed-dotted line)}
 \label{fig:spat_dist}
 \end{figure*}

\section{Cluster member selection}
\label{sec:gal_sel}
\begin{figure} 
 \centering 
  \resizebox{\hsize}{!}{\includegraphics{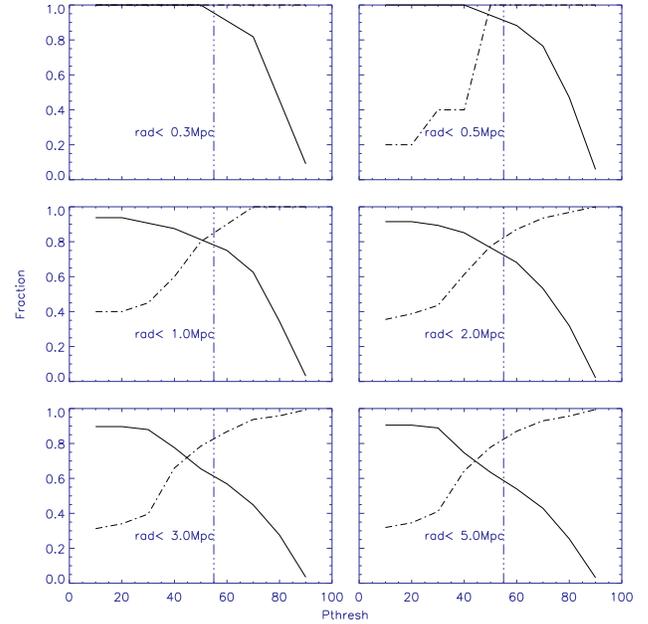}}
 \caption{Galaxy cluster member selection accuracy for MS 0451.5-0305 . The different panels show the results for different radii. Solid lines indicate the completeness and dashed lines the contaminations from non-members as a function of the probability threshold. The vertical dashed line shows the adopted probability threshold.} 
 \label{fig:gal_member}
 \end{figure}

\begin{figure} 
 \centering 
  \resizebox{\hsize}{!}{\includegraphics{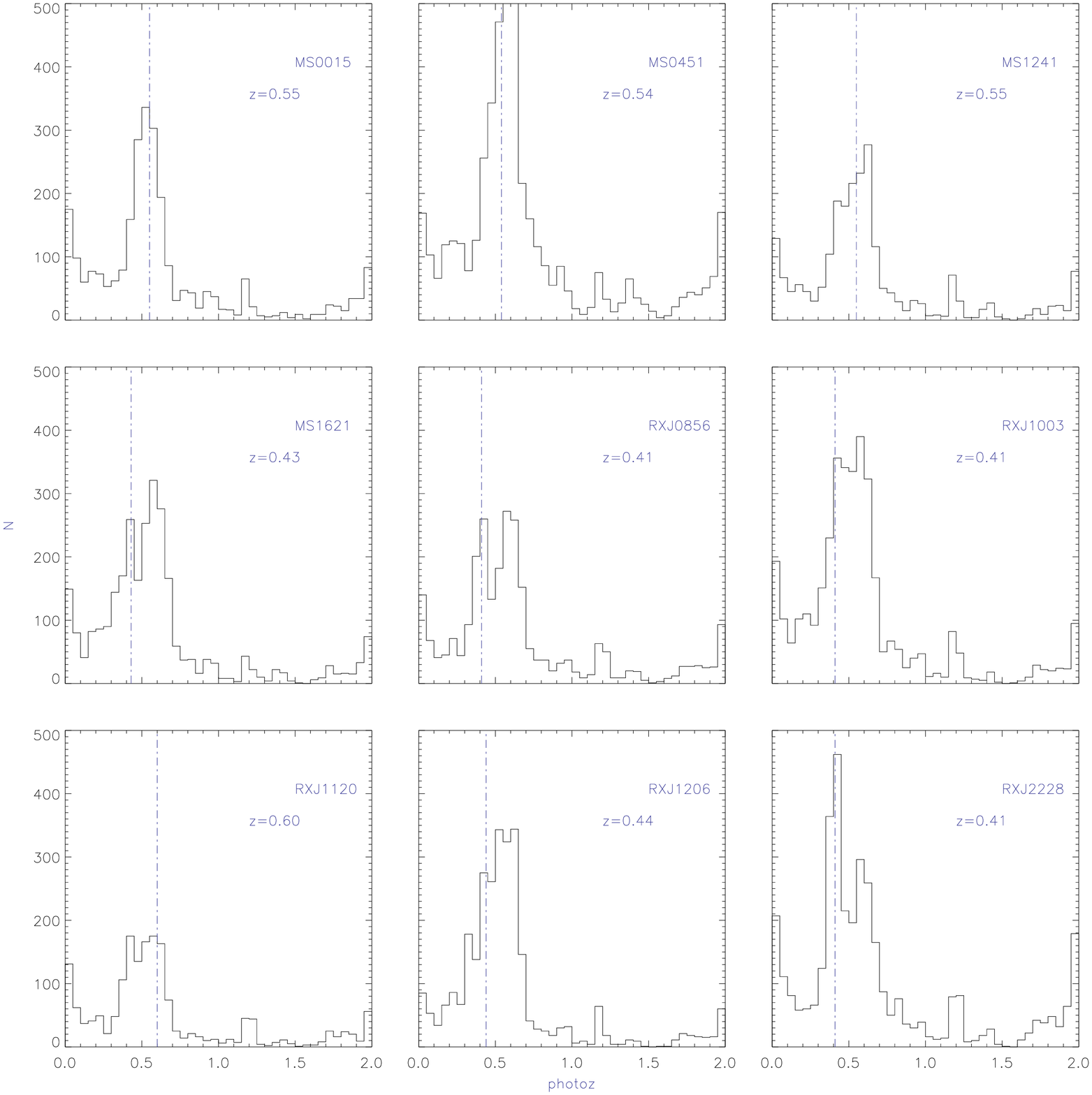}}
 \caption{Photometric redshift distributions for all the clusters. Dashed vertical lines show the cluster spectroscopic redshift. } 
 \label{fig:zs_vs_zp}
 \end{figure}

The most unambiguous way to determine cluster membership is by means of accurate spectroscopic
redshifts. Unfortunately, it is too time consuming to obtain high-sepctroscopic completeness,
even to relatively bright limits. He we adopt a method for membership determination based on
photometric redshift estimates presented in \cite{Pello09}. Usually, $z_{phot}$-based methods
for determining cluster membership are based on a simple cut in redshift, such that a galaxy
is considered to be member if $|z_{phot}-z_{cluster}| < \Delta z_{thresh}$. The problem of
this kind of approach is that $z_{thresh}$ can be as high as 0.3, producing high field
contaminations. As a matter of fact, the accuracy of photometric redshifts as a function of redshift strongly depends on the filter set and its ability to detect strong spectral features, such as the 4000$\AA$ break. In this case, the contiguous spectral coverage from g to z is well suited to derive good-quality photometric redshifts for galaxies between $0.2<z<1.2$, i.e. within the redshift range of interest for our study. Galaxies below $z\sim0.2$ or beyond $z\sim1.2$ will have poor-quality redshifts and broad redshift probability distributions.  We therefore expect this population to contaminate the $z_{phot}$ distribution in these regions (see Fig.~\ref{fig:zs_vs_zp}). To reduce this contamination, we combine both the best-fit redshift and the
information contained in the full redshift probability distribution $P(z)$, by computing the integrated probability for the galaxy to be in the cluster ($P_{clust}$):
\begin{equation}
P_{clust}=\int^{z_{cluster}+0.1}_{z_{cluster}-0.1}P(z)dz
\end{equation}
Since the contaminating galaxies 
do not exhibit strong features within the griz spectral domain (i.e. their broad redshift-probability distributions avoid the cluster redshift), this selection criterion based on the probability distribution should remove most of them. 
By using a representative spectroscopic sample, which is available for MS0451.5-0305
\citep{Ellingson98} we can calibrate the threshold value $P_{thresh}$ above which 
a galaxy is considered to be a cluster member. This criteria is then extrapolated to the rest
of the
sample, since all the clusters lie in the same redshift range and are observed in similar
conditions (same detector, integration time and set of filters).  
We apply this method to all the objects detected with \textsc{SExtractor} within $r<5$ Mpc and
$18<r<23$. We use the \textsc{HyperZ} code \citep{Bolzonella00} and the 4 available
photometric bands to compute photometric redshifts (Fig.~\ref{fig:zs_vs_zp}). For color indices, we use the MAG\_APER
magnitudes, which give the flux on a fixed $2.5$ arcsec aperture, corrected from zeropoint
errors. As there remains some uncertaintites in the photometric calibration in some filters,
we add some positive or negative offsets to the zeropoints of the different filters determined
so that the color-color relations of the stars detected in the images fit those of the Pickles library \citep{Pickles98}. \\
Figure~\ref{fig:gal_member} shows the evolution of the rejection of field galaxies as well as
the completeness of the members selection for different values of $P_{thresh}$ and for
different radii. We clearly see that increasing the probability threshold results in an
increase of the fraction of rejected non-members while the fraction of cluster members
progressively decreases. Since our main objective is to study the effects of the cluster
environment on morphology and color we want to reduce the number of field contaminations. We
therefore consider as galaxy cluster members all the objects which have
$|z_{phot}-z_{cluster}|<0.1$ and $P_{cluster}>55\%$. As shown in Fig.~\ref{fig:gal_member},
this ensures at least $80\%$ completeness and $\sim80\%$ rejections up to $r<2$Mpc. Above this
distance, the completeness is slightly lower ($\sim70\%$) but the rejection remains high. We
establish the limit of 5 Mpc as the distance at which there is still a reasonable compromise
between completeness and field contamination, well above the virial radius of the clusters. This results in a sample of 3861 galaxies distributed in the 9 clusters. We show in figure~\ref{fig:spat_dist}, the spatial distribution of galaxies in the 9 clusters. 

\section{Physical properties}
\label{sec:phys_prop}

\subsection{Morphology}
\label{sec:morpho}
\subsubsection{Method}
We now discuss the morphological classification of the objects belonging to the cluster. Morphologies are determined in the r-band images using \textsc{galSVM}\footnote{\url{http://www.lesia.obspm.fr/~huertas/galsvm.html}}. \textsc{galSVM} is a non-parametric N-dimensional code based on support vector machines (SVM) that uses a training set built from a local visually classified sample. \citep{Huertas-Company08a}. Basically we separate galaxies in two broad morphological types (late and early type).  By late-type we mean spiral and irregular galaxies (S/I) and early-type galaxies include elliptical and lenticular types (E/S0). \\
Broadly, the employed procedure can be summarized in 4 main steps (see \citealp{Huertas-Company08a} for more details):
\begin{enumerate}
\item Build a training set: we select a nearby visually classified sample at wavelengths corresponding to the rest-frame of the high redshift sample to be analyzed to minimize morphological k-correction effects. In our case, we want to simulate r-band observations at $z\sim0.5$. We use therefore an SDSS local sample observed in the g band which roughly corresponds to the rest-frame wavelength. We then move the sample to the proper redshift and
image quality, assign a magnitude randomly selected in the magnitude range [19:23] with a
distribution representative of the observed luminosity function and drop it in a real image to properly simulate the background noise.

\item Measure a set of morphological parameters on the sample. In this particular case, the classification is made in 7 dimensions, including: ellipticity, asymmetry, gini, concentration, $M_{20}$ and smoothness. For more details on how these parameters are exactly computed please consult \cite{Huertas-Company08a} 

\item Train a support vector based learning machine with a fraction of the simulated sample and use the other fraction to test and estimate errors.

\item Classify real data with the trained machine.
\end{enumerate}

\subsection{Accuracy of classifications: cross-checking with HST}

One of the main challenges of the present work is performing a morphological classification on seeing-limited images. Until now, all the works involving morphological classifications of distant galaxy clusters have been based on HST imaging. We need consequently to carefully investigate the accuracy of our classification. Fortunately, two of our clusters have published data obtained with the space telescope. This is the case of MS 1621.5+2640 \citep{Ellingson97, Saintonge05} and MS 0451.5-0305  \citep{Ellingson98,Moran07}. We therefore have a reference to calibrate our morphological classification on seeing-limited images.

\begin{figure} 
 \centering 
  \resizebox{\hsize}{!}{\includegraphics{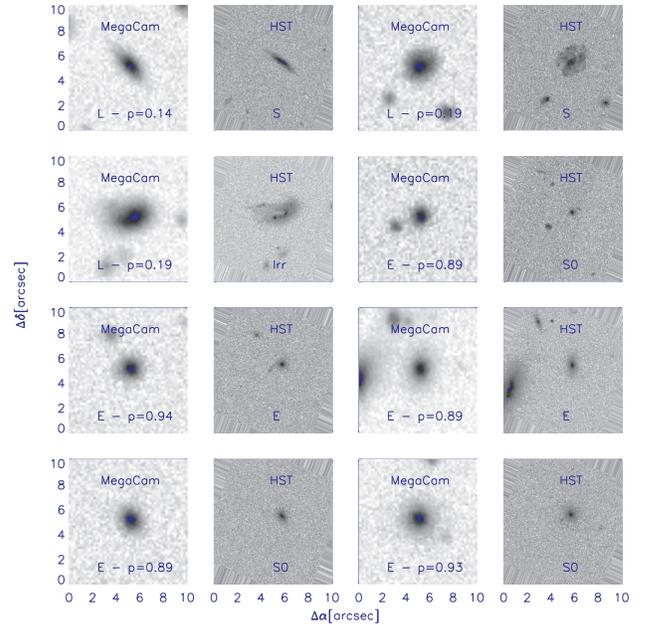}}
 \caption{Stamps of the same galaxies seen by HST and MegaCam. The stamps size is $10^{"}\times10^{"}$. For each galaxy, we indicate the morphological class obtained from the MegaCam image (L=late-type, E=early-type) as well as the probability of being early-type and the visual classification of \cite{Moran07} for the HST stamps. } 
 \label{fig:morpho_stamps} 
 \end{figure}

\begin{figure} 
 \centering 
  \resizebox{\hsize}{!}{\includegraphics{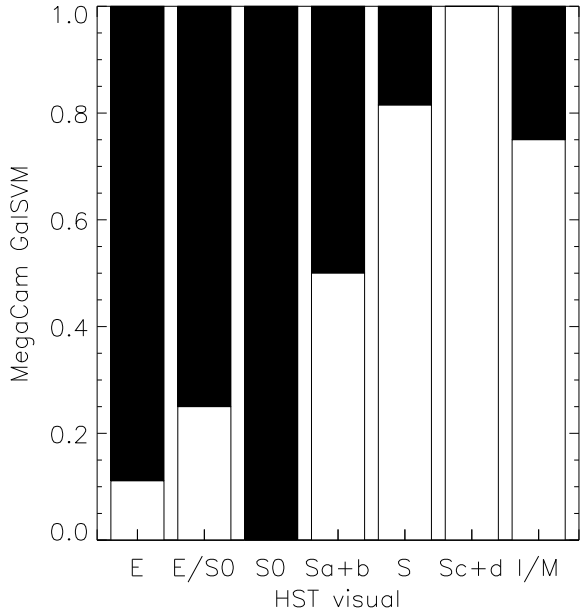}}
 \caption{Relation between the visual morphology computed in \cite{Moran07} and the \textsc{galSVM} type. Filled histograms show the fraction of \textsc{galSVM} early-type objects and empty histograms the fraction of late-type galaxies.} 
 \label{fig:MS0451_HST_galSVM} 
 \end{figure}

\paragraph{MS 0451.5-0305 } This cluster has available published results on 96 galaxies based on HST imaging which enables a second calibration of our classification. \cite{Moran07} performed detailed visual morphological classifications using WFPC2 imaging. Figure~\ref{fig:MS0451_HST_galSVM} shows the comparison of our \textsc{galSVM}-based classification with their results. Both classifications are in good agreement: a big fraction ($\sim90\%$, 19/21) of early-type objects according to HST (i.e. E +S0) fall in the \textsc{galSVM} early-type class and 33/41 late-type galaxies (i.e. S+Sc+Sd+I) fall in our late-type class. Most of the \emph{problems} therefore seem to come from the intermediate morphology class. Indeed, we are not able to separate properly the Sa/b galaxies which seem to randomly fall in one or the other class. It implies then that we are slightly over-estimating the early-type fraction. 
For illustration, we show in figure~\ref{fig:morpho_stamps} some stamps of the same galaxies seen by HST \citep{Moran07} and MegaCam. 

\paragraph{MS 1621.5+2640} \cite{Saintonge05} performed an automated morphological
classification based on model-fitting of 258 galaxies. They used the bulge fraction ($B/T$) as
the main parameter to split objects into different morphological classes. In
table~\ref{tbl:BT_prob} we compare the probability of being early-type according to our
classification and the bulge fraction of \cite{Saintonge05} for the 59 galaxies identified as
cluster members and within our selected magnitude range:  a galaxy with a low bulge fraction
should have also a low probability of begin elliptical. Of course the dispersion of the distribution is considerable but we have to keep in mind that we are not measuring the same parameters, so points do not necessarily have to lie on the unity slope line. Since we are only interested in splitting the galaxies into two rough morphological types, we have to look at the fraction of galaxies that are clearly miss-classified, if we consider the HST-based classification as a reference (which is not necessarily true since it is an automated classification as well). This includes objects which are clear disc-dominated galaxies ($B/T<0.3$) but have a probability of being early-type greater that 0.5 and vice-versa,  galaxies with a bulge fraction bigger than 0.7 but low probabilities of being ellipticals. We find that the first represent 5 objects out of 41 early-type objects and the second one out of 18. We conclude then that our classification is robust and in good agreement with the HST one. Note however that we might be slightly over-estimating the fraction of elliptical galaxies if compared with HST studies as pointed out in the previous comparison. If we indeed consider galaxies with an intermediate bulge-fraction (i.e. $0.3<B/T<0.7$), most of them (25/31) fall in the early-type class. It worth noticing here that despite of these differences with HST, this work presents 2 important advantages: first, the size of the sample ($\sim4000$ galaxies) makes it more robust to miss-classifcations and second, we are using the same method to classify galaxies up to $5$ Mpc, in all density regions, which implies that any difference between the center and the outskirts must be real. 

\begin{table} 
 \centering 

\begin{tabular}{lcc|}
\hline\hline\noalign{\smallskip}
 & $p_{early}>0.5$ & $p_{early}<0.5$ \\
\noalign{\smallskip}\hline

$B/T>0.7$ & 30\% ( 11)& 5\% (1)  \\ 
$B/T<0.3$ & 10\% ( 5 ) & 62\%  (11) \\ 
$0.3<B/T<0.7$ & 60\% ( 25 ) & 33\% (6) \\ 
\noalign{\smallskip}\hline
\noalign{\smallskip}\hline\noalign{\smallskip}

\end{tabular}

 \caption{Relation between the bulge fraction computed in \cite{Saintonge05} and the probability of being early-type computed by \textsc{galSVM}.} 
 \label{tbl:BT_prob}
 \end{table}

\subsection{Stellar masses}
\begin{figure*} 
 \centering 
  \resizebox{\hsize}{!}{\includegraphics{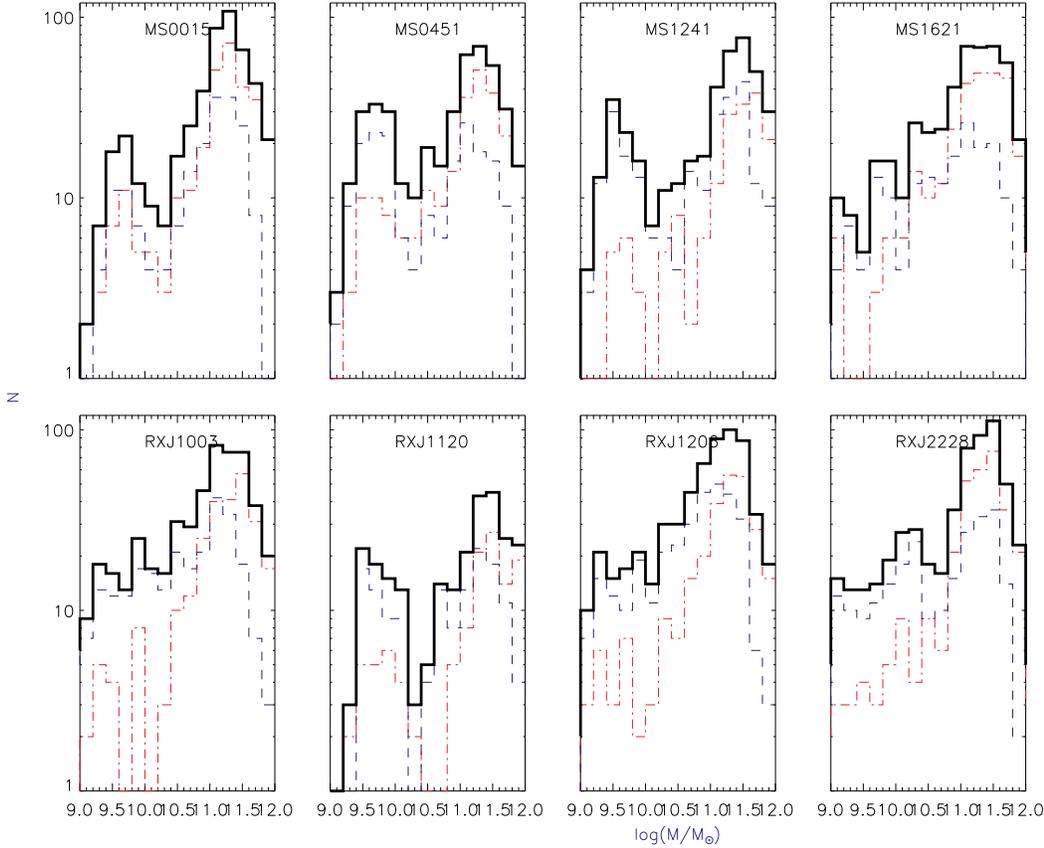}}
 \caption{Stellar mass histograms for all the galaxies in the 9 clusters. Solid black lines show the histogram for all the galaxies while color lines show the histograms separated by morphological type (blue: late-type, red: early-type).} 
 \label{fig:mass_dists} 
 \end{figure*}

Recent works have shown that the relations between galaxies and their environment in clusters strongly depend on the considered mass range \citep{Patel09, Holden07}. \cite{Holden07} did not find, for example, any evolution on the early-type population with $M>10^{10.6}$ solar masses, using a mass-limited sample. They suggested therefore that most of the morphological transformation which occurs from $z\sim0.8$ must affect the low mass galaxies. It appears therefore interesting to investigate the effect of stellar mass on the different analyzed trends in our sample. In this work, stellar masses are computed directly from the best-fitting model of the photometric redshift estimate using a \cite{chabrier03} IMF. The mass estimate takes into account the age of the model as well as the star forming history. Even though we do not have NIR photometry, we consider our mass estimates reliable since at $z\sim0.5$ the redder filters (i,z) are probing the SED around $5000\AA$, (beyond the $4000\AA$ break) where the flux is mostly coming from evolved stellar populations. \\
Mass distributions for all the clusters are shown in figure~\ref{fig:mass_dists}. They are clearly bimodal for most of the clusters, with late-type galaxies dominating the low-mass regime ($log(M/M_{\odot})<10.5$) and the early-type ones dominating the high-mass one $log(M/M_{\odot})>11$, with of course some cluster-to-cluster variations.
 We try to estimate the mass completeness of our sample using the magnitude cut for
the morphological classification ($r'<23$), which is about 2 magnitudes brighter than the
magnitude limit of the catalogs. If we compare the mass function of the morphologically
classified sample with that of the full sample up to the  magnitude limit ($r'=25$), we find that it is
complete at the 50\%\ level for a mass of $\log(M/M_\odot)=9.5$ (or M $\simeq 2.5 \ 10^{9}
M_\odot$). In the following, we will therefore divide the sample in three stellar mass bins: $9.5<log(M/M_\odot)<10.3$, $10.3<log(M/M_\odot)<11.3$, $log(M/M_\odot)>11.3$. While the two last bins are not affected by incompleteness, the less massive one can be, specially for the early-type population which is mostly red and is therefore lost earlier in a magnitude limited survey. Nevertheless, we have decided to include this mass bin since some information can be extracted anyway. However, we are aware that this might introduce some biases when interpreting our results so we will properly point out and take into account this point when necessary in the following sections. Note also that a rough estimate of the apparent magnitude of a $L^\star$ galaxy at
$z \sim 0.5$ is $m^\star_r = 20.6$ so the magnitude limit for the morphological classification
corresponds to a $L^\star /10$ galaxy. 
 
\subsection{Local density}

\begin{figure} 
 \centering 
  \resizebox{\hsize}{!}{\includegraphics{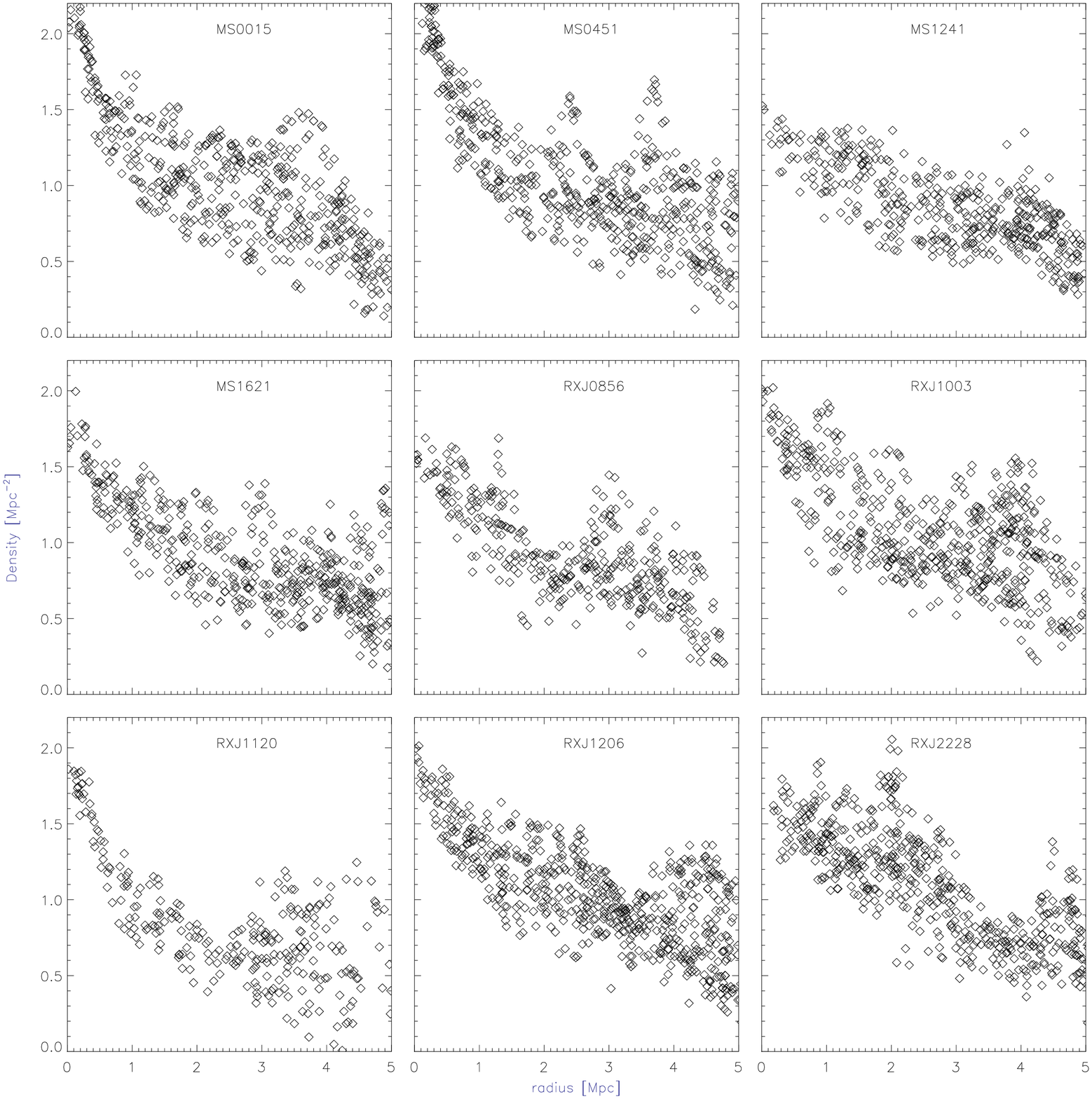}}
 \caption{Local density vs. radius for all the clusters in our sample.} 
 \label{fig:dens_rad} 
 \end{figure}

The local density has been shown to play a key role in the evolution of galaxies, since the efficiency of the different mechanisms involved in such transformation often depends on the local density. To characterize it for all the galaxies in our sample we compute the local projected galaxy density, $\Sigma$, at each galaxy location. 
We use the n-th nearest neighbor definition, with n=10, to compute $\Sigma$: 
\begin{equation}
\Sigma=\frac{n+1}{\pi d_n^{2}}
\end{equation}
where $d_n$ is the projected distance of the n-th nearest neighbor. The value of $n=10$ is commonly used in clusters since it is a good measure of the local over-density in this high density environments \citep{cooper05}. 
We include in the computation, all the cluster members selected with the method presented in \S~\ref{sec:gal_sel}. To avoid edge biases, we include galaxies a bit further of our $5Mpc$ distance limit.
We show in figure~\ref{fig:dens_rad}, the relationship between radius and local density. We clearly see that local density is not a well-defined function of radius, except in the central $\sim1$ Mpc. We confirm then that by studying the cluster outskirts we are able to isolate the effects of over-density from the radius , and deduce which one of both effects most influences the morphological evolution of galaxies (\S~\ref{sec:morpho_dens}). Departures from spherical symmetry in the distribution of galaxies in the
clusters will be enhanced by using the local density instead of the radial distance to the
cluster center. This is more prominent in cluster outskirts, where anisotropies of the cluster
distribution or the presence of in-falling groups are better taken into account. That way, we
can better evaluate how local effects influence the morphological evolution of galaxies (\S~\ref{sec:morpho_dens}).

\section{Results and discussion}
\label{sec:results}
\subsection{Composition of the red-sequence}
\label{sec:RS}
Mass-color diagrams are a powerful tool to study the star-formation (SF) activity of galaxies. Galaxy distribution in this plane is bimodal, with passive red galaxies lying in a tight mass-color relation called the red-sequence (RS) and blue-star forming galaxies presenting a wider range of colors and masses and therefore lying in the so-called blue-cloud. This bimodality is usually
analysed through the color-magnitude relation (e.g. \citealp{Mei09}). The advantage of the 
color-stellar mass relation is that it displays a less scattered relation, especially for the blue
cloud. This is can be understood because there is a correlation between color and M/L ration
for late-type galaxies which is partly removed in a stellar mass-color relation
and gives a narrower blue cloud (e.g. \citealp{Patel09}). 
In terms of galaxy evolution in dense environments, as galaxies enter in the cluster, we
expect that their star-formation processes are 
quenched by the different interaction effects. Galaxies move progressively 
from the blue-cloud to the red-sequence, while suffering a morphological
transformation from late-types to earlier types. By looking at the composition of the
red-sequence we can identify which galaxies have already suffered the transformation in
clusters at $z\sim0.5$ and then try to isolate the processes responsible of this
transformation. Nevertheless, this scheme is too simple to interpret all the diversity of galaxy
morphologies and cannot explain the status of the "red late-type galaxies" or the "blue
early-type" ones, but at least part of the bimodality is imprinted by evolution processes.



 \begin{figure*}
 \centering 
  \resizebox{\hsize}{!}{\includegraphics{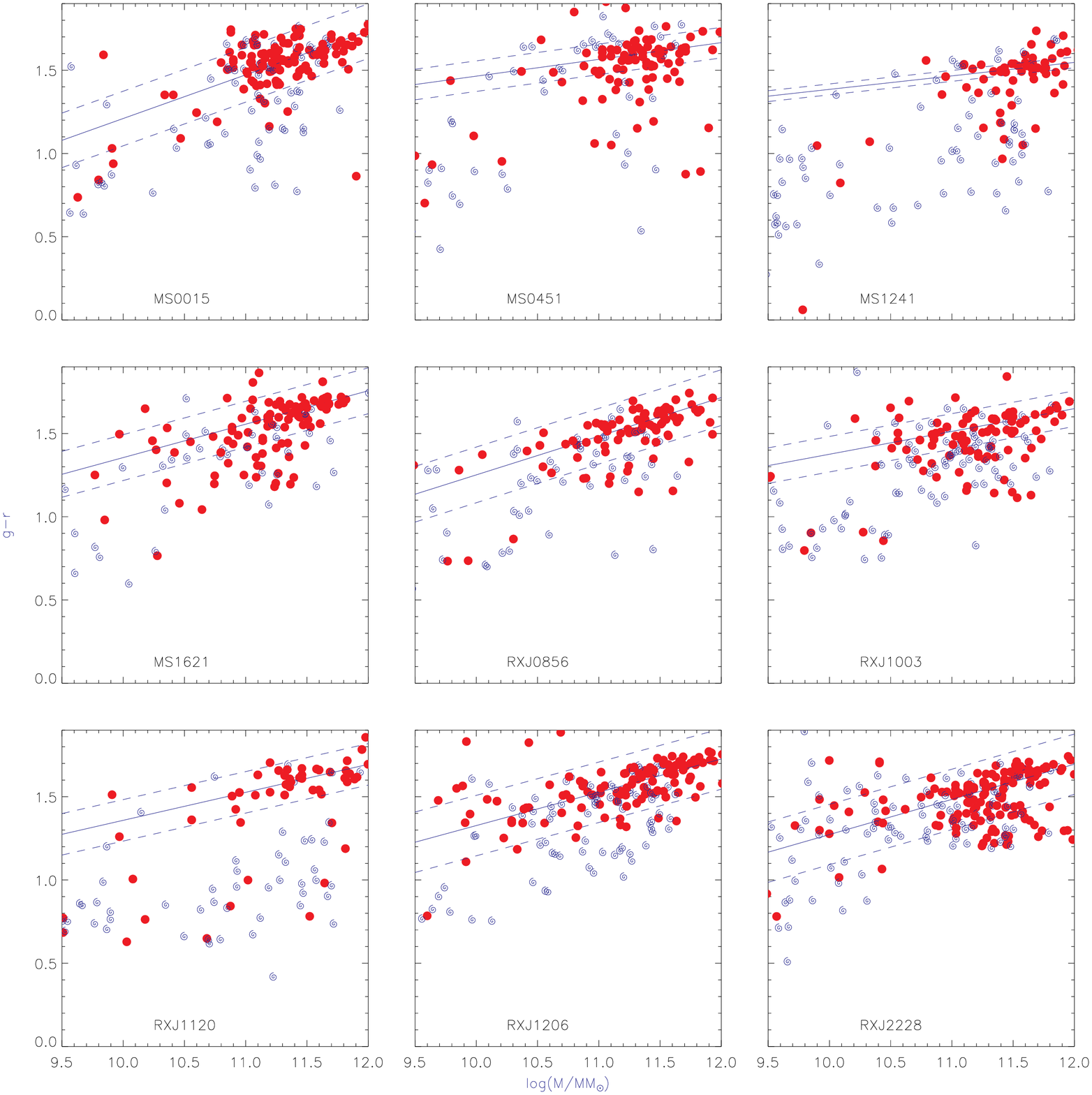}}
 \caption{Stellar mass-color plots in the 2 central Mpc for all the 9 clusters in the sample. Early-type galaxies are shown with red filled circles and late-type with blue spiral symbols. Solid line indicate the best fit to the red-sequence and dashed lines $1.5\sigma$ limits (see text for details). } 
 \label{fig:mag_color} 
 \end{figure*}

Figure~\ref{fig:mag_color} shows the g-r-stellar mass diagrams in the central 2 Mpc of all the clusters in our sample. We define whether a galaxy belongs or not to the red-sequence based on its deviation from the color-mass relation (CMR) shown in figure~\ref{fig:mag_color} (black solid line). The CMR is derived by performing an iterative fit to early-type galaxies with $log(M/M\odot)>10.5$ and rejecting the $2\sigma$ outliers of the best-fit at each iteration upon convergence. A galaxy is then said to belong to the red-sequence if it lies within the $2\sigma$ scatter of the non-rejected galaxies (dashed lines). It is worth noticing that despite of the fact that we see the same trends in all the clusters, we find that the CMR is not universal and that it is subjected to a cluster-to-cluster variance over the cluster population with different scatters and abundances. It is beyond the scope of this paper to analyze in details the shapes of the red-sequences but we are planning to do it in a coming paper. However, some interesting information can be extracted from these plots. First, they help checking the mass completeness limit of our sample and verifying that red objects (mostly early-type) are indeed lost earlier. As a matter of fact, above $log(M/M_\odot)\sim10.3$ we do clearly see the RS while below this mass, red galaxies start to become less abundant. This fact confirms that results obtained in the low mass bin ($9.5<log(M/M_\odot)<10.3$) might be affected by incompleteness and should be analyzed carefully. 
Second, it is possible to study the general composition of the red-sequence. If we first look at the morphological distributions, we clearly find out the expected trend, i.e. the majority of object lying in the RS are early-type objects (76\%, table~\ref{tbl:morpho_red}). This result just confirms that as galaxies enter in the cluster and transform into elliptical galaxies, they progressively become redder and move to the red-sequence. \\
Furthermore, the majority of early-type objects lying the red sequence are massive objects ($log(M/M_{\odot})>11.3$) and as we progressively move into lower masses the number of galaxies of later morphological types becomes more important. 

These tendancies with the stellar mass suggest that most of the evolution in these intermediate redshift clusters is affecting low mass galaxies ($log(M/M_{\odot})<11.3$), while the most massive galaxies seem to be well in place in the red-sequence. These results seem to be in agreement with \cite{delucia07} who also found that, at intermediate redshift, luminous galaxies are already in place in the RS and evolve passively, while there is a deficit of faint red galaxies, suggesting that these galaxies are populating the RS from the blue cloud between $z\sim0.8$ and today. Furthermore, semi-analytical models (e.g. \citealp{Menci08}) also predict that at a cosmic time of $t\sim2.5$ Gyr ($z\sim2.5$), more than $90\%$ of the stellar mass finally assembled in red-sequence galaxies at $z\sim1$ is already in place.

\begin{table*}
\caption{Composition of the red sequence in the 2 central MegaParsecs. Table shows the fraction of galaxies of each morphological and mass bins which belong to the red sequence.  }
\begin{tabular}{lccccc|}
\hline\hline\noalign{\smallskip}
& $log(M/M_\odot)>11.3$  & $10.3<log(M/M_\odot)<11.3$ & $9.5<log(M/M_\odot)<10.3$ & Total\\
\noalign{\smallskip}\hline
E/S0 & 45\% & 27\% & 4\% & 76\%\\
S/I  & 8\% & 12\% & 4\% & 24\%\\
\noalign{\smallskip}\hline
\noalign{\smallskip}\hline\noalign{\smallskip}
\label{tbl:morpho_red}
\end{tabular}
\end{table*}

\subsection{Radial trends}
\label{sec:morpho_rad}
\begin{figure*}
 \centering 
  \resizebox{\hsize}{!}{\includegraphics{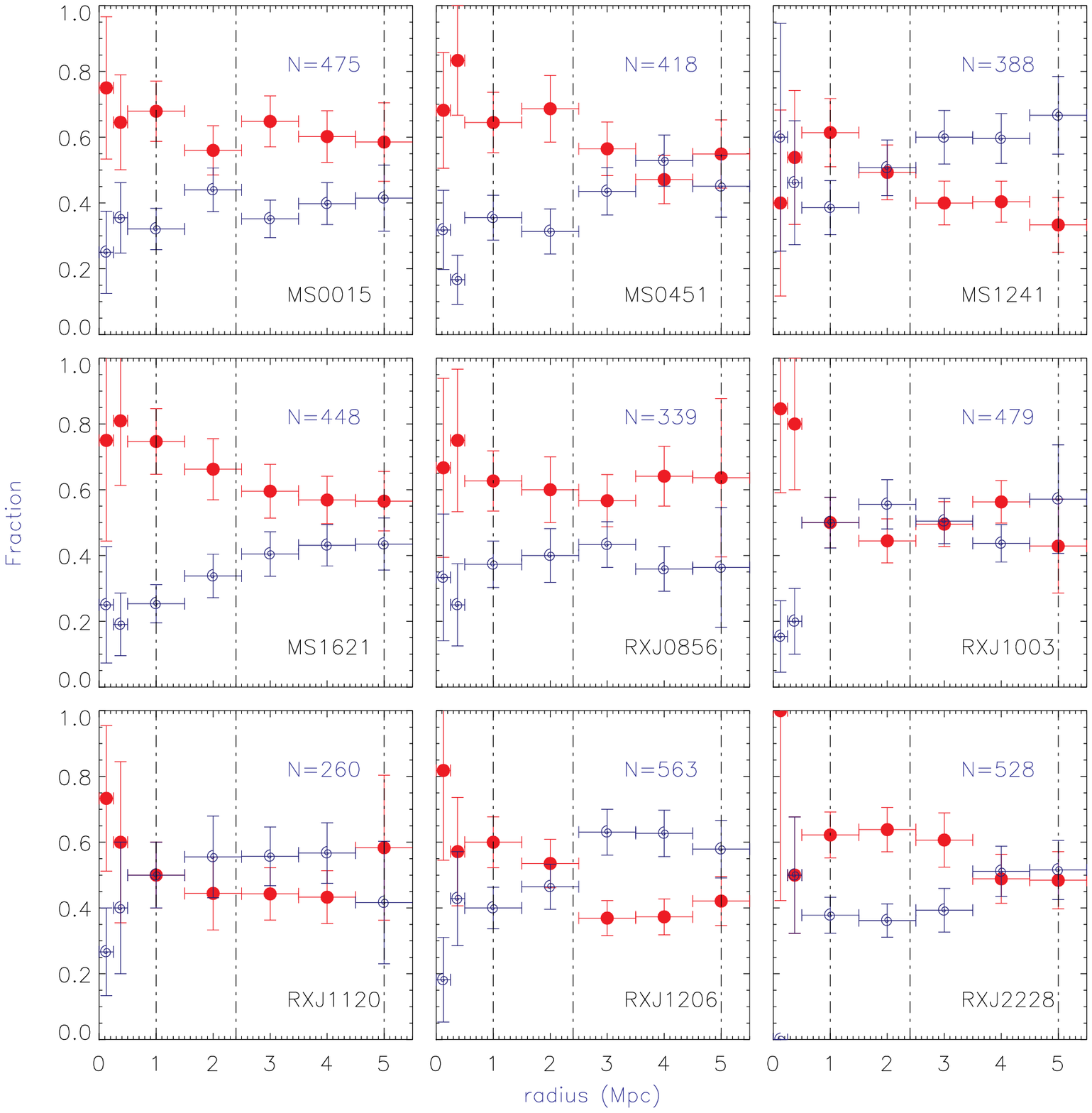}}
 \caption{Fraction of morphological types ($r<23$) as a function of cluster radius. Red points: E/S0 galaxies, blue points: S/I galaxies. The total number of objects is shown in each plot.} 
 \label{fig:morpho_rad} 
 \end{figure*}

The morphologies of galaxies are correlated with distance from the cluster center in nearby clusters (e.g. \citealp{Whitmore93}) and groups (e.g. \citealp{Postman84}) as well as at higher redshifts (e.g. \citealp{Abraham96, vanDokkum00}). An increase of the fraction of early-type galaxies towards the center is observed in all cases, but the detailed evolution of the morphology-radius relation from high redshift remains uncertain. Figure~\ref{fig:morpho_rad} shows the fractions of early and late-type galaxies at different radii for clusters in our sample. All the clusters present similar trends: as detected in previous works, we observe a correlation between morphology and distance, with elliptical galaxies representing about $80\%$ of the galactic populations in the inner parts ($d<0.5$ Mpc) and decreasing at larger distances. More precisely, we see a steep decrease of the early-type fraction in the central Megaparsec which goes from $\sim80\%$ to $\sim60\%$ and then a mild decrease in the outer parts from $\sim60\%$ to $\sim40\%$ in agreement with previous HST-based works (e.g.~\citealp{Treu03}). These two different regimes are normally explained by different processes. While the explanation of the steep increase of the early-type fraction in the inner cores requires phenomena related to the cluster potential, such as tidal interaction or ram-pressure stripping which accelerate the morphological transformation, this is not the case in the outskirts. As a matter of fact, in the periphery, most of the galaxies have never been through the center and therefore are free from effects such as ram-pressure stripping. The mild decrease in the outskirts might be explained by a progressive quenching of the star formation (starvation, \citealp{Balogh00}) or after a starburst phase (e.g. \citealp{Poggianti99}). \\

Another possibility is that the morphology radii relation is the consequence of a segregation effect: objects with different radii represent galaxies with different early assembly histories (through mergers) that keep a correlation between their history and the location in the cluster (e.g. \citealp{Diaferio01}). 

\begin{figure*}
 \centering 
  \resizebox{\hsize}{!}{\includegraphics{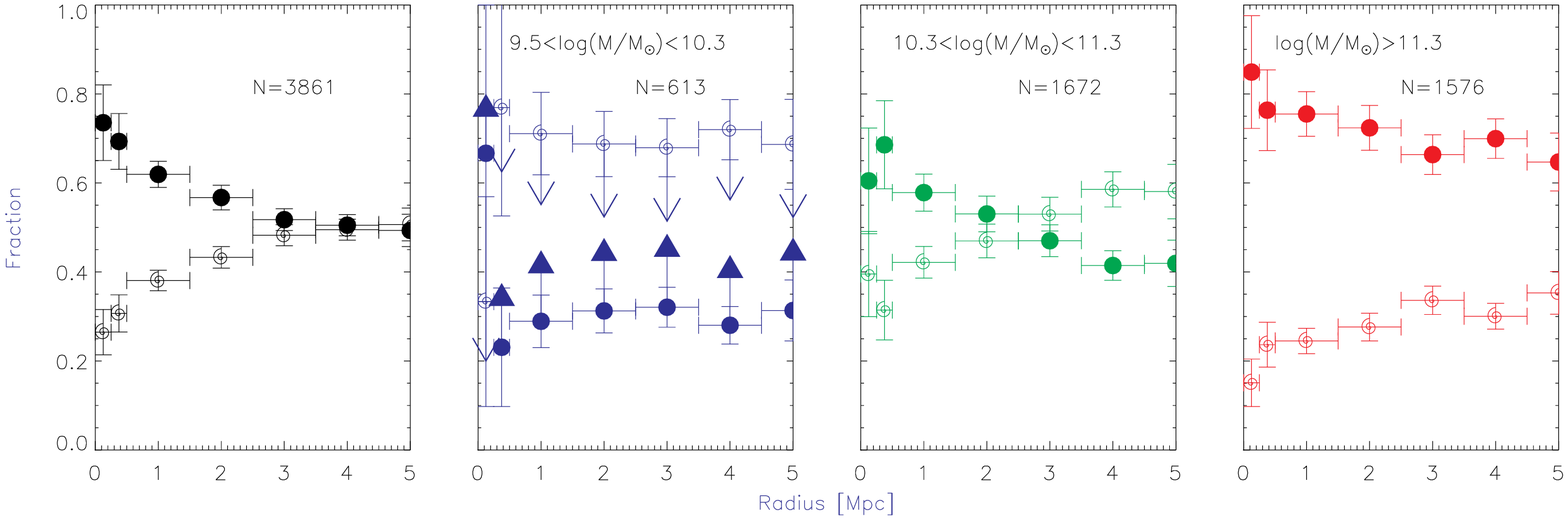}}
 \caption{Morphological mixing as a function of the cluster radius, averaged for all the clusters in the sample and for different mass ranges. Filled circles are early-type galaxies and spiral symbols are late-type galaxies. The left panel shows the results for the whole sample. Notice that for the low mass bin (top right) the two first points are not significative since there are too few objects. The total number of objects is shown in each plot. In the low mass bin, vertical arrows show an estimate of the possible effects of incompleteness (see text for details).}
 \label{fig:morpho_rad_avg} 
 \end{figure*}

If this last scenario reveals to be true massive galaxies should have had their star formation activity decrease faster than less massive galaxies and therefore should be already predominantly early-type when they enter the cluster at $z\sim0.5$. Figure~\ref{fig:morpho_rad_avg} shows the radial trends for different mass ranges, averaged for all the clusters in order to have better statistics. The fraction of massive galaxies ($log(M/M_{\odot})>11.3$) remains almost constant at all radii, indeed suggesting that these galaxies were formed before they entered the cluster, probably in infalling groups through galaxy-galaxy interactions and consequently experience small evolution. As a matter of fact, evidence that the source of the constituent galaxies of massive clusters are groups in the outskirts has also been seen in previous works \citep{Dressler97, Gonzalez05, Moran07, Patel09}. The presence of these galaxies in the cluster center can perhaps be explained by a segregation effect. \\
The general trend of a decrease of early-type galaxies with radius has to come therefore from lower mass galaxies,  especially the intermediate mass range ($10.3<log(M/M_{\odot})>11.3$). Most of the evolution in the cluster galaxy population therefore seems to be occurring in this stellar mass range. On the contrary, for the 
lowest mass bin in our sample ($9.5<\log(M/M_{\odot})<10.3$), the galaxy population 
is dominated by late-type galaxies as if they did not have time to evolve yet. Note that we do
not consider the very central part of the cluster ($d < 0.5$ Mpc) because there are too few
low mass galaxies and numbers are not significant enough. We have to be however very cautious when analyzing this low mass bin, since as stated in~\S \ref{sec:RS}, it might be strongly  affected by incompleteness, leading to an under-estimate of the early-type population. As a mater of fact, since most of the galaxies lost in this mass bin are probably red early-type galaxies, the values measured should be considered as lower limits. In order to give a rough estimate of the maximum value, we considered that the 50\% of galaxies are E/S0s and estimated how the fraction of these galaxies would rise at every position (see Fig.~\ref{fig:morpho_rad_avg}). We observe that, even if we loose half the early-type population at all distances, it would remain below the late-type population.
 

\subsection{Morphology-density relation}
\label{sec:morpho_dens}
\begin{figure} 
 \centering 
  \resizebox{\hsize}{!}{\includegraphics{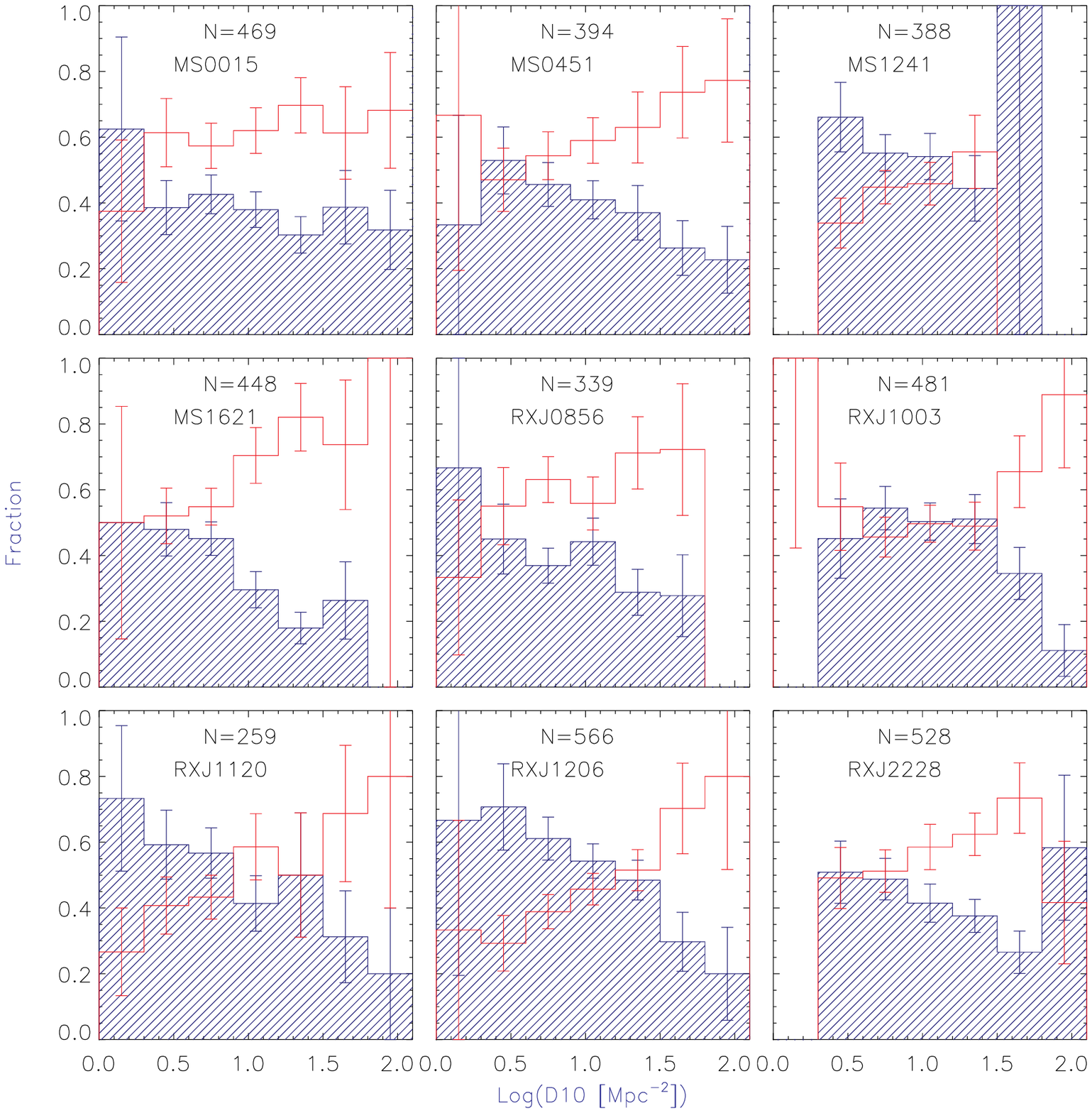}}
 \caption{Morphology-density relation for all the clusters in the sample. Blue lines are late-type systems and red lines are early-type. The total number of objects is shown in each plot. } 
 \label{fig:morpho_dens} 
 \end{figure}

Figure~\ref{fig:morpho_dens} shows the $T-\Sigma$ relation for all the clusters in the sample. In all clusters, the fraction of early-type galaxies increases monotonically with local surface density, with some heterogeneity due to cluster-to-cluster variations. This relation
is not entirely correlated with the decrease of the early-type fraction with radius (the $T-R$
relation) even if there are some similarities. In practice, we see that, beyond $\sim1$Mpc,
the correlation between radius and density (Fig.~\ref{fig:dens_rad}) is weaker with
a high dispersion in the number density so that we span more different environments. 

\begin{figure}
 \centering 
  \resizebox{\hsize}{!}{\includegraphics{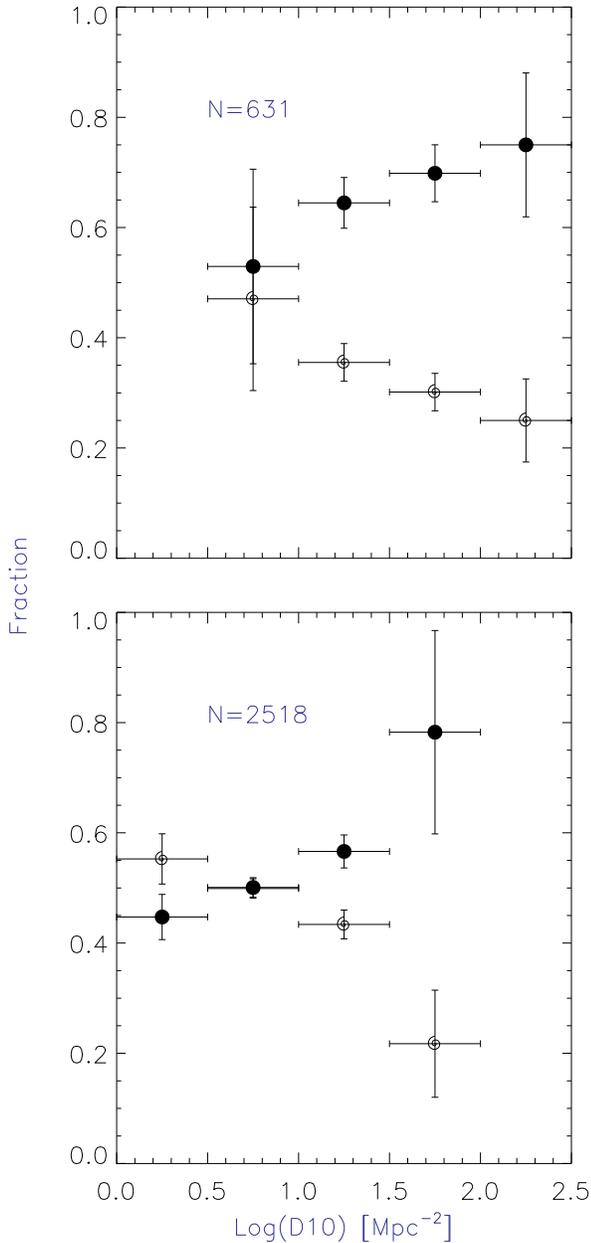}}
 \caption{Mean fractions of early-type (filled circles) and late-type (spiral symbols) galaxies as a function of local density in the cluster center (top panels) and in the cluster outskirts (bottom panels). The total number of objects is shown in each plot.}
 \label{fig:morpho_dens_mass_avg_total} 
 \end{figure}

\begin{figure*}
 \centering 
  \resizebox{\hsize}{!}{\includegraphics{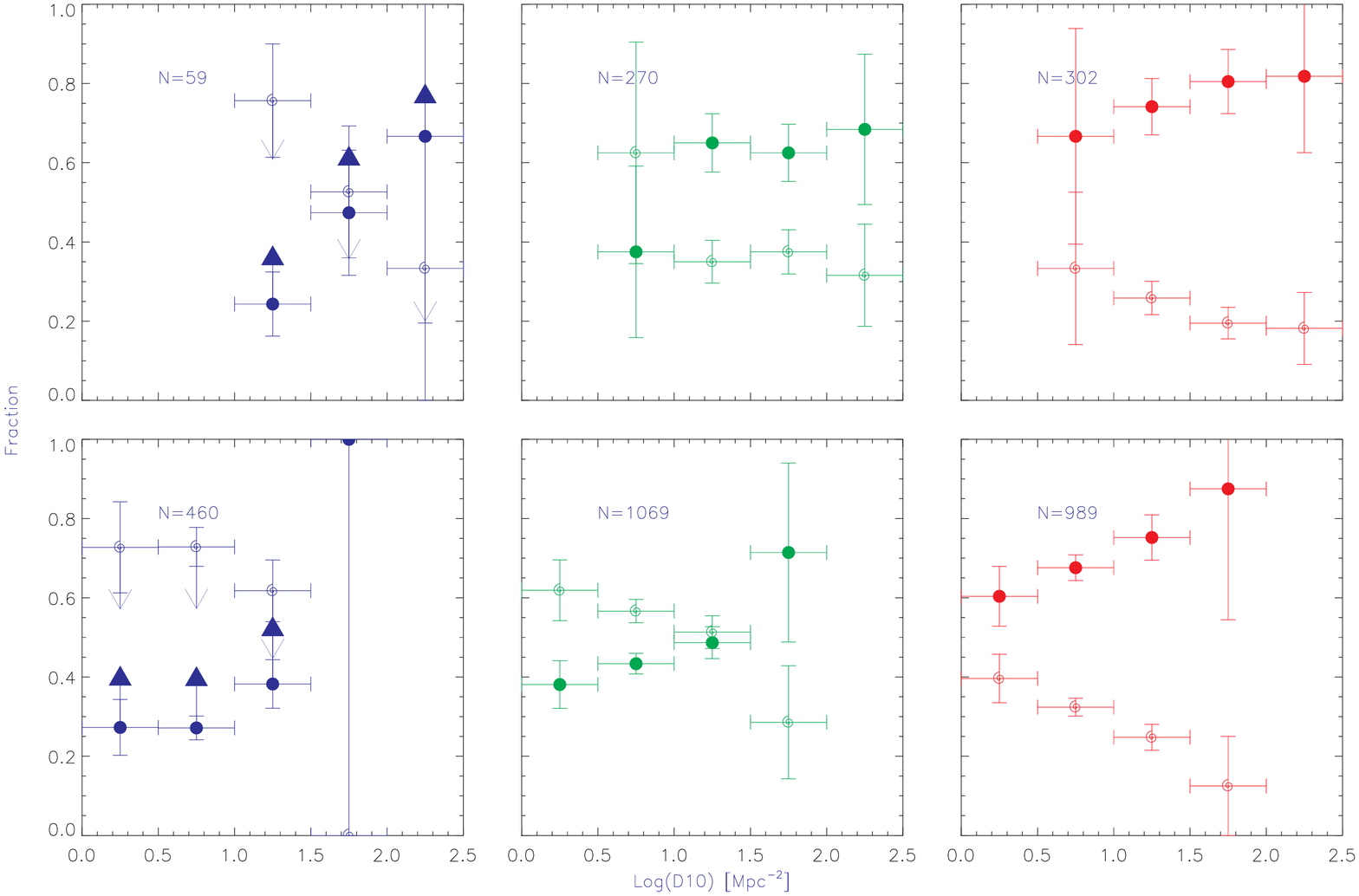}}
 \caption{Mean fractions of early-type (filled circles) and late-type (spiral symbols) galaxies as a function of local density in the cluster center (top panels) and in the cluster outskirts (bottom panels) and for different mass ranges. Colors indicate three different mass ranges: red ($log(M_{*}/M_{\odot})>11.3$), green ($10.3<log(M_{*}/M_{\odot})<11.3$) and blue ($9.5<log(M_{*}/M_{\odot})<10.3$). The total number of objects is shown in each plot. In the low mass bin, vertical arrows show an estimate of the possible effects of incompleteness (see text for details).}
 \label{fig:morpho_dens_mass_avg} 
 \end{figure*}

Therefore, as shown in previous works, studying the properties of galaxies in the center ($d<1Mpc$) and in the outskirts ($d>2Mpc$) separately, helps separating different physical phenomena. While at the center, the effects of local environment (density) and cluster potential (distance to the center) are very well correlated, this is not the case in the outskirts. Consequently, differentiating the analysis of galaxy properties in the
outskirts with respect to radial distance $R$ or local number density $\Sigma$ will tell us if
galaxies are entering the cluster in groups and evolving afterwards or if they are acquiring
the relation between morphology and environment/radius once they enter the cluster. \\
Figure~\ref{fig:morpho_dens_mass_avg_total} indeed shows the trend with density of the early-type fraction averaged over all the clusters for galaxies in the outer parts ($d>2Mpc$) and in the central regions ($d<1Mpc$). The fraction of early-type galaxies is systematically higher in all the environments in the center than in the outskirts. This rise of the global
fraction of E/S0 objects reflects the fact that galaxies are experiencing a strong
morphological segregation in cluster centers. In the outskirts of the cluster, the global
morphological mixing is correlated with local density, i.e. the fraction of early-type
galaxies rises with increasing density and the $T-\Sigma$ relation is preserved. This seems to
indicate that morphological transformation of galaxies entering the cluster is strongly
influenced by the local environment and more precisely by interactions with galaxies in their
close surroundings. The correlation is also present in the cluster center, but with a higher fraction of E/S0 galaxies at all redshifts. Galaxies therefore continue evolving inside the cluster and migrating to the red-sequence but keeping a memory of their surroundings. \\
Are all galaxies evolving at the same speed and independantly of their mass? In figure~\ref{fig:morpho_dens_mass_avg} the fractions of each morphological type for three stellar mass ranges in the center and in the outskirts of the cluster are shown. We detect different behaviors depending on the considered mass range. Massive galaxies $log(M_{*}/M_{\odot})>11.3$ are mostly elliptical ($\sim70\%$) in the cluster outskirts at all the environments. The fraction is slightly lower than the one measured in the cluster center ($\sim80\%$) but remains high. This seems to indicate that an important fraction of
galaxies in this mass range are already formed when they enter the cluster at $z\sim0.5$ and
have already experienced morphological evolution in their local environment (tidal interactions and merging in groups, \citealp{bamford09}). \\
Intermediate mass objects ($10.3<log(M_{*}/M_{\odot})<11.3$) show a different trend. If we consider the cluster outskirts, we do see a relation between the morphological mixing and the local density: galaxies in the higher density bin ($log(D10)>2.0$) are mostly early-type ($\sim60\%$) while the fraction decreases to $\sim40\%$ at lower densities. This population is therefore experiencing an evolution driven by the local environment when entering the cluster and dominates the general trend seen in the whole sample (fig.~\ref{fig:morpho_dens_mass_avg_total}). If we now look at the central regions, the correlation is almost absent and the fraction of early-type galaxies rises in all the density bins. This again suggests that this population is evolving as it moves to the cluster center but galaxies still preserve a memory of their location when they entered the cluster.\\
If we consider now low massive galaxies ($9.5<log(M_{*}/M_{\odot})<10.3$), we again detect a different behavior. First, this population is dominated by late-type objects at almost all densities, even if we still measure a higher fraction of E/S0 galaxies at high densities, i.e. the cross-over between late-type and early-type systems occurs at higher density than for the more massive galaxies. It seems that morphological transformation is still going on even close to the cluster center \citep{Holden07}. Once again, this result might be a consequence of incompleteness in this mass bin, which leads to an under-estimate of the red early-type population. As for section~\ref{sec:morpho_rad} we show with vertical arrows in figure~\ref{fig:morpho_dens_mass_avg} how the fractions would increase if we consider that we loose 50\% of the early-type population. Taking into account this effect reduces the trend, but there are still more late-type galaxies at all densities than than in the more massive bins. \\
Therefore, the transition from the blue-cloud to the red sequence is performed at different speeds mainly driven by the local environment: more massive galaxies drift to the red-sequence at earlier epochs. Massive galaxies have converted the gas into stars earlier probably because at high redshift their star formation was not effectively self-limited by feedback. Most of these galaxies are already massive, gas-poor and with an early-type morphology when they enter the cluster. 
The transition for intermediate/low-mass galaxies is still going-on at $z\sim0.5$ and seems to be mainly driven by the local environment as stated by the correlation between morphology and density at all the cluster locations. Starbusts resulting from galaxy merging or starvation processes can be responsible of this transition: galaxy encounters can significantly accelerate the star formation and provoke the transition to the red-sequence. Semi-analytical models show however that this process is effective mainly for midsize galaxies  which strike the best compromise between cross section and abundance \citep{maulbetsch07}. Galaxy interactions can also trigger the AGN phase. The AGN feedback can then contribute to suppress any residual star formation. 

\section{Summary and conclusions}

We have presented a morphological study of a sample of galaxies in 9 X-ray luminous clusters
at intermediate redshift ($z \sim 0.5$). Wide field imaging provides information up to 5 Mpc
from the cluster center, well above the virial radius and allows to explore the distant
outskirts of the clusters, where the local environment is dominant with respect to cluster
distance. Using photometric redshifts we are able to select cluster members with less than
20\% of field  contamination, thanks to a detailed comparison with spectroscopic redshifts
when available. Galaxies are then separated into two morphological classes with our automatic
galaxy classifier \textsc{galSVM}, developped and validated by \citet{Huertas-Company08a}.
Direct comparison with HST imaging of a sub-sample of galaxies in two independant clusters
demonstrates that our method, based on ground-based data, is in agreement with HST results for
more than 90\% of the galaxies up to a magnitude limit $r'=23$. This magnitude cut which
defines our working sample is about 1.5 mag. fainter that a $L^\star$ galaxy at $z \sim 0.5$
and ensures multi-color photometric completeness. In addition, multi-color photometric information,
available in 4 bands ($g', r', i'$ and $z'$), enables an estimate the stellar mass of the
galaxies from their SED fitting. We estimate that our sample is 50\% complete down to a
stellar mass of $3 \ 10^9 \ M_\odot$ (or $\log(M_{*}/M_{\odot}) \sim 9.5$. All in all, our
full galaxy sample contains 3861 galaxies inside 9 clusters. \\
The main results can be summarized as follows:
\begin{itemize}
\item In all clusters, we clearly identify the red sequence of early-type galaxies and we are
able to build a stellar mass versus color sequence of galaxies. Both sequences are mostly
populated with massive ($\log(M_{*}/M_{\odot})>11.3$) early-type galaxies. As found by
previous studies, the galaxy red sequence is already formed at $z \sim 0.5$ and early-type
galaxies properties are already well in place \citep{Patel09}. Physical processes
which occur in dense environments have consequently a rapid and early influence of the
evolution of galaxy properties (harassment, ram pressure strippring, etc ...). They are very
efficient in cluster centers and operate early in the cluster histories so that galaxies
evolve quickly towards gas poor massive early-type galaxies.

\item The population of massive galaxies ($\log(M_{*}/M_{\odot})>11.3$) does not seem to have
evolved significantly since $z \sim 0.5$. In fact, the fraction of early-type galaxies among
this population is very high and remains almost constant ($T-R$ relation), from the cluster
center ($\sim90\%$) to the outskirts ($\sim70\%$ at the virial radius and above), suggesting
that these galaxies experience no evolution. They were probably formed at higher redshifts
before entering the cluster in infalling groups and their location in the cluster center might be mostly explained by a segregation effect.


\item When we consider the whole sample, the $T-\Sigma$ relation shows a clear correlation
between morphology and local density with early-type galaxies dominating the densest regions.
This was initially pointed out by \citep{Dressler80} in the local universe and confirmed later
on more distant clusters (e.g. \citealp{Postman05}). 
The $T-\Sigma$ relation is also clearly detected in
the cluster outskirts, where cluster radius and local density are no longer correlated. It
seems therefore that the morphological evolution is mainly driven by the local environment
through galaxy-galaxy interactions (merging, starvation), independently of cluster
properties. Galaxies entering the cluster keep a memory of their surroundings and evolve
accordingly. They evolve however at different speeds, depending on the stellar mass. 

\item Most of the evolution within clusters at $z\sim0.5$ is taking place in galaxies with
intermediate mass ($10.3<\log(M_{*}/M_{\odot})<11.3$) or even lower. These galaxies do show a
relation between the morphological mixing and the local density in both the center and the
outskirts with differences between dense and loose global environments. In dense environments,
early-type galaxies are in place in a similar way as for the most massive ones, while in
external areas, the $T-\Sigma$ relation is similar to the one in the field. So the
morphological dichotomy is prominent in structures representative of galaxy groups but in
poorer environments, the morphological transformation from spiral-types to early-types has not
produced significant changes in the galaxy distribution. Galaxies in this mass range are experiencing a migration from the blue-cloud to the red-sequence as they move into the cluster, mainly driven by interactions with their surroundings. The migration is accelerated when galaxies, within their local environment, cross the densest parts of the clusters in the central Mpc or so. 

 \item Galaxies in the smallest mass bin ($9.5<\log(M_{*}/M_{\odot})<10.3$) are basically
late-type galaxies at all the clusters locations and environments (even if they present a
trend with the local density), suggesting that they do no have started the morphological
transformation at $z\sim0.5$. However, our conclusions are weaker for this population because
the number of galaxies is small, especially in the cluster center and results can be affected by incompleteness specially in the early-type population. However, even considering that $50\%$ of early-type galaxies are lost in this mass bin, this population seems to be dominated by late-type galaxies. 

\end{itemize}

The picture which seems to arise from all these results is that massive galaxies moved to the red-sequence well before $z\sim0.5$. Lower mass galaxies ($log(M_{*}/M_{\odot})<11.3$) are still in process of migration from the blue-cloud to the red-sequence. The migration is mainly driven by galaxy-galaxy interactions which cause a progressive quench of the star formation and/or provoke strong starbusts and/or trigger an AGN phase. Only in the central Mpc or so, the cluster potential seems to play an important role by accelerating the process of migration. 

In order to better trace the general scheme of the morphological evolution of galaxies in
different environments, two approaches are necessary. First, we wish to analyse in more
details the dependence of these global tendancies with the general properties of the clusters
themselves. This will be done in a future work as soon as the mass distribution of the
clusters is available using weak lensing mass maps (work in progress, Foex et al. in
preparation) and/or X-ray properties such as luminosity and temperature.
Clusters of galaxies are not a fully homogeneous population and some
of the evolution properties of their cluster members are attached to their intrinsic
caracteristics \citep{Poggianti09}. 
Second, it is also necessary to extend this approach to more distant clusters to connect clusters at different epochs and trace an evolutionary track. In order to
better probe the stellar mass and the morphology traced by the old stellar population,
near-infrared imaging is preferred \citep{Huertas-Company09}. 


\begin{acknowledgements}
We are grateful to C. Lidman and M. West for many
  interactions and helpful discussions. The authors want to thank as well the anonymous referee for useful comments which clearly helped to improve the paper.
  We also thank the Programme National de
  Cosmologie of the CNRS and ESO for financial support. 
  \end{acknowledgements}

\bibliographystyle{aa}
\bibliography{biblio}

\end{document}